\documentstyle[psfig]{mn}
\newif\ifAMStwofonts
\newcommand{\lapp}{\mbox{\raisebox{-0.3em}{$\stackrel{\textstyle <}{\sim}$}}}

\title[A spiral-host episodic radio galaxy]
      {Discovery of a spiral-host episodic radio-galaxy}
\author[Ananda Hota et al.]
    {Ananda Hota$^1$,$\thanks{E-mail: hotaananda@gmail.com, hota@asiaa.sinica.edu.tw}$ 
S.K. Sirothia$^2$, Youichi Ohyama$^1$, C. Konar$^1$, Suk Kim$^3$,
\newauthor Soo-Chang Rey$^3$, D.J. Saikia$^2$, J.H. Croston$^4$ and Satoki Matsushita$^1$  \\
$1$ Academia Sinica Institute of Astronomy and Astrophysics, P.O. Box 23-141, Taipei 106, Taiwan, R.O.C. \\
$2$ National Centre for Radio Astrophysics, TIFR, Post Bag 3, Ganeshkhind, Pune 411 007, India \\
$3$ Department of Astronomy and Space Science, Chungnam National University, Daejeon 305-764, Republic of Korea\\
$4$ School of Physics and Astronomy, University of Southampton, Southampton, SO17 1BJ, U.K.\\
}
\date{Accepted.    Received}
\pagerange{\pageref{firstpage}--\pageref{lastpage}}
\pubyear{  }
\begin{document}
\maketitle
\begin{abstract}
We report the discovery of a unique radio galaxy at $z$=0.137, which
could possibly be the second spiral-host large radio galaxy and also
the second triple-double episodic radio galaxy. The host galaxy shows
signs of recent star formation in the UV but is optically red and is
the brightest galaxy of a possible cluster. The outer relic radio
lobes of this galaxy, separated by $\sim$1 Mpc, show evidence of
spectral flattening and a high fraction of linear polarisation.  We
interpret that these relic lobes have experienced re-acceleration of
particles and compression of the magnetic field due to shocks in the
cluster outskirts.  From the morphology of the relics and galaxy
distribution, we argue that re-acceleration is unlikely to be due to a
cluster-cluster merger and speculate about the possibility of
accretion shocks.  The source was identified from SDSS, GALEX, NVSS
and FIRST survey data but we also present follow up optical
observations with the Lulin telescope and 325 MHz low frequency radio
observations with the GMRT. We briefly discuss the scientific
potential of this example in understanding the evolution of galaxies
and clusters by accretion, mergers, star formation, and AGN feedback.
\end{abstract}
\section{Introduction}
\begin{figure*}
\hbox{
  \psfig{file=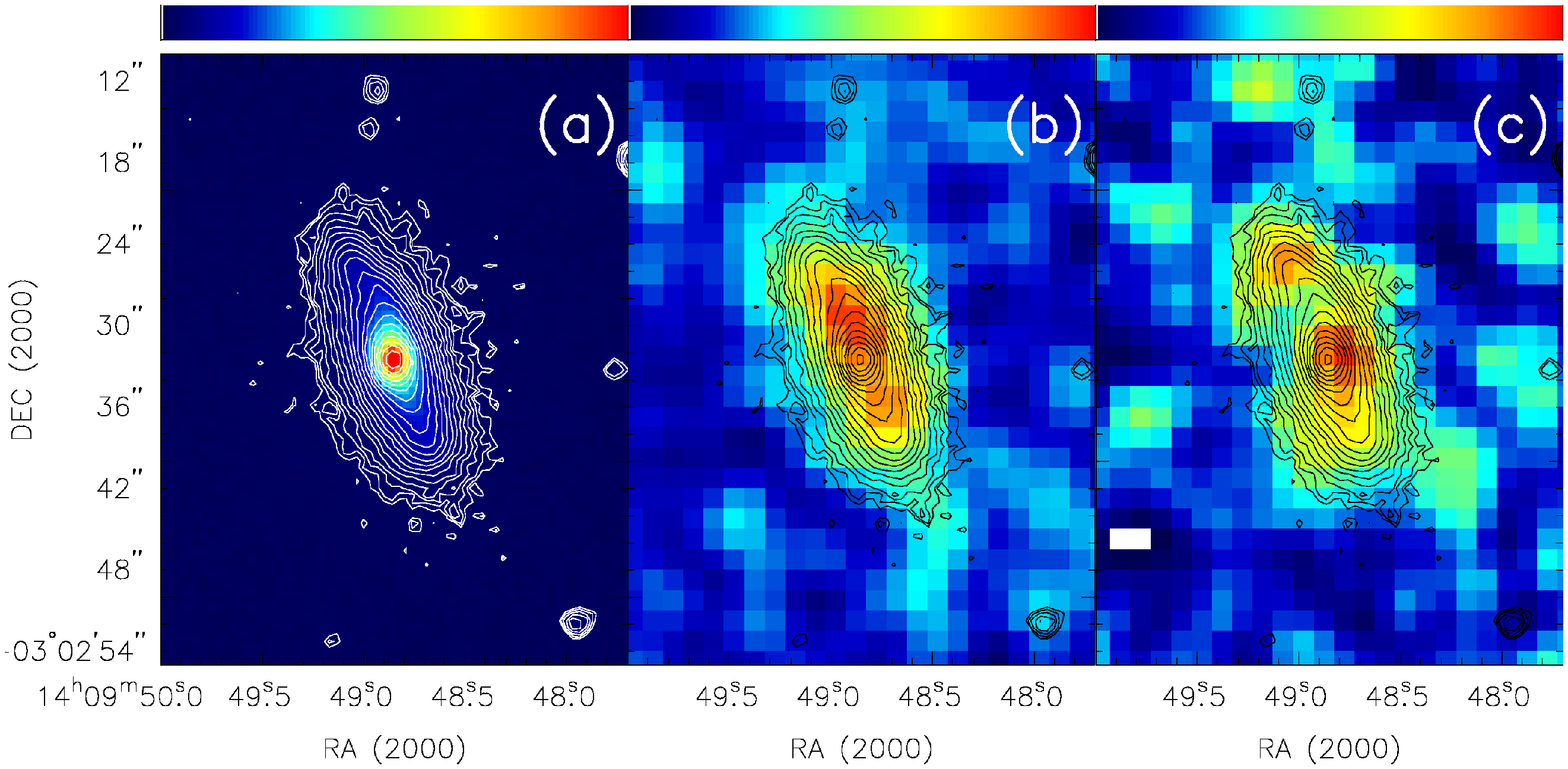,width=4.1in,angle=-90}
   \psfig{file=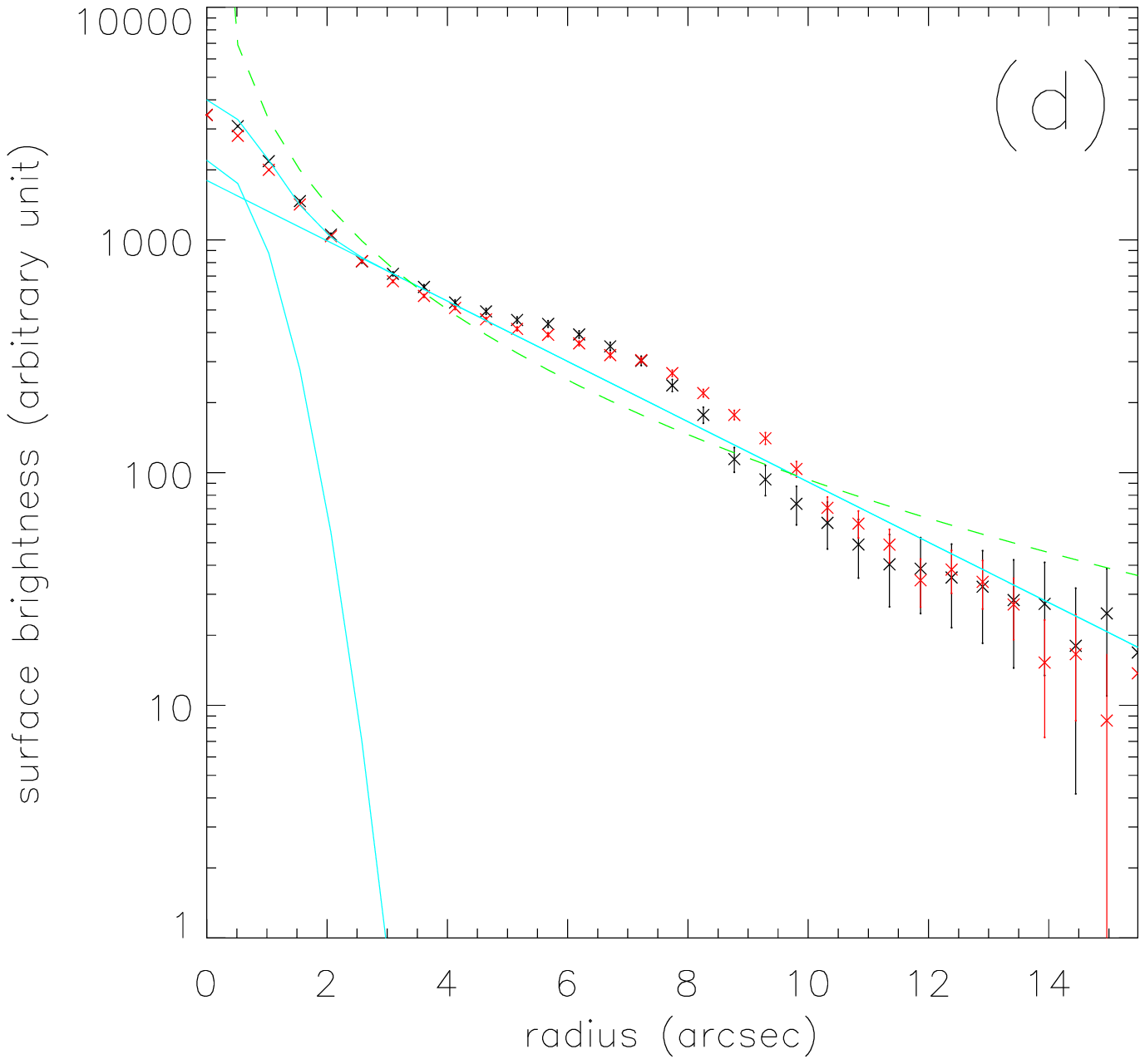,width=2.6in,angle=0}
  }
\caption[]{(a) Lulin $R$-band image of the host galaxy Speca, presented both in colour (linear scale) 
and contours (logarithmic scale).
(b) The same $R$-band contours are superposed on the smoothed near-UV image from GALEX in linear scale.
(c) The same $R$-band contours are superposed on the smoothed far-UV image also from GALEX in linear scale.
(d) Radial surface brightness profile of the Lulin $R$-band image.
Black and red data are shown for the SW and NE parts of the disk, respectively, for comparison.
Fitted model profiles are of a nuclear point source and exponential disk (cyan continuous line) and of a de Vaucouleurs profile (green broken line).
For the former model, both individual model components and the total are shown.
              }
\end{figure*}
By incorporating processes of AGN feedback into the models of
hierarchical structure formation, it is possible to explain observed
galaxy population properties, such as the galaxy and cluster
luminosity function, the $M_\bullet$$-\sigma$ relation, and the
colour-magnitude bi-modality of nearby galaxies (Springel, Di Matteo
\& Hernquist 2005, Croton et al. 2006).  Powerful and large-scale
feedback in the form of radio lobes seen as multiple cavities in X-ray
images of clusters has provided strong support for such models (Nulsen
et al. 2005, Sanders \& Fabian 2007, Randall et al. 2011).  Luminous
radio galaxies are also found to be tracers of high-$z$ cluster
formation (Venemans et al. 2007, Zirm et al. 2005).  In the nearby
Universe, diffuse Mpc-scale shell-like radio emission found in the
outskirts of clusters has been interpreted as emission from
acceleration of particles at shock fronts due to cluster mergers or
structure formation (Ensslin et al. 1998; Ensslin et al. 2001; Bagchi
et al. 2006, 2011; van Weeren et al. 2010).  

Apart from mergers, clusters grow by accretion from cosmic galaxy
filaments. Evidence of shocks, predicted for such accretion, is yet to
be convincingly detected from X-ray observations, but many radio
observations show promising prospects (Ensslin et al. 1998, Pfrommer,
Ensslin \& Springel 2008).  A wide-angle tail radio galaxy recently
discovered to be present in a several-Mpc-scale galaxy filament has
been proposed as evidence of gas accretion through the filament onto
the cluster (Edwards et al. 2010).  Due to the Mpc-scale size and
episodic nature of radio galaxies (Saikia \& Jamrozy 2009), whose
relic plasma as old as 2 Gyr can be revived (Ensslin \& Gopal-Krishna
2001), low frequency radio observations, naturally tracing old
relativistic plasma, are powerful probes of these merger, accretion,
and feedback processes driving cluster formation and evolution.  The
time-scale of 1$-$2 Gyr is comparable to the dynamical timescale of
galaxies in the clusters, the timescale of galaxy mergers, the
timescale of the star formation in UV-bright early-type galaxies (Yi
et al. 2005) and quenching of star formation in the post-starburst
(E+A) galaxies (Goto 2005).  Therefore, combined low-frequency study
of relics of AGN feedback and recent star formation in AGN host
galaxies is a potential new strategy of investigating galaxy-black
hole co-evolution via mergers and feedback.

Here we report the discovery of a unique radio galaxy whose host galaxy 
appears to be a spiral or disk with young star formation, rather than a typical 
giant elliptical galaxy with a predominantly old stellar population. 
Equally unique is that it shows signs of three distinct episodes of jet ejection where the 
outer relic radio lobes, separated by $\sim$1 Mpc, have experienced particle 
re-acceleration possibly due to shocks in the cluster outskirts. 
The optical source had been cataloged before as SDSS J140948.85-030232.5 or 
LCRS B140713.3-024824 or MaxBCG J212.45357-03.04237 (being brightest in the cluster) 
but for ease in referring we nicknamed it, due to its uniqueness, 
`Speca' (SPiral-host Episodic Cluster-dominant AGN).
For the spectroscopic redshift of Speca $z$=0.1378, 1 arcsec corresponds to  2.396 kpc,
in a Universe with $H_0$=71 km s$^{-1}$ Mpc$^{-1}$, $\Omega_{\rm m}$=0.27 and
$\Omega_{\rm vac}$=0.73.
\section{Observations and data analysis}
{\bf Optical imaging:} Optical $R$-band (Bessell) observations were
made on the night of 2010 May 16 (UT) with the Lulin One-meter
telescope (LOT).  A CCD camera (PI1300B;
Kinoshita et al. 2005) with an EEV CCD36-40 chip was attached at the
Cassegrain focus of the telescope to provide a pixel scale of 0.51
arcsec.  We took 17 frames of 180 sec exposures, and they were
processed in a standard manner (dark and bias subtraction, flat
fielding with a sky flat, stacking with sigma clipping for cosmic-ray
removal).  The final stacked image of 3060 sec exposure has an
effective (after stacking) seeing size of 1.5 arcsec (FWHM). \\ 
{\bf Radio continuum imaging:} The galaxy was observed with the Giant
Metrewave Radio Telescope (GMRT; Swarup et al. 1991) at 325 MHz on
2010 April 26 with a total bandwidth of 32 MHz and total time on the
target galaxy $\sim$5 hours.  The data reduction was done mainly using
{\tt AIPS++}~(version: 1.9, build \#1556).  3C48 was the primary flux
density and bandpass calibrator. After applying bandpass corrections
on the phase calibrator (1419+064), gain and phase variations were
estimated, and then flux density, bandpass, gain and phase calibration
from flux density and phase calibrators were applied on the target
field.  The standard procedure for radio continuum data analysis was
followed and a final image was made after several rounds of phase
self-calibration, and one round of amplitude self-calibration, where
the data were normalized by the median gain for all the data.  The
final images resulted in an r.m.s. noise of 0.2 mJy beam$^{-1}$ and
angular resolution of 9.75 arcsec $\times$ 7.45 arcsec (Position Angle
(P.A.)=63$^\circ$). The low-resolution map prepared to match the beam
size (45 arcsec) of NRAO VLA Sky Survey (NVSS)
has a higher noise (1.4 mJy beam$^{-1}$).  All the radio images presented
in this letter have uniform leveling of contours, r.m.s noise $\times$
(-4, -2.82, 2.82, 4, 5.65, 8... in steps of $\sqrt{2}$).
\section{Results and Discussion}
\begin{figure}
\vbox{
  \psfig{file=SPECA-GMRT-ACTIVE-SE-RELIC.PS,width=2.5in,angle=0}
  \psfig{file=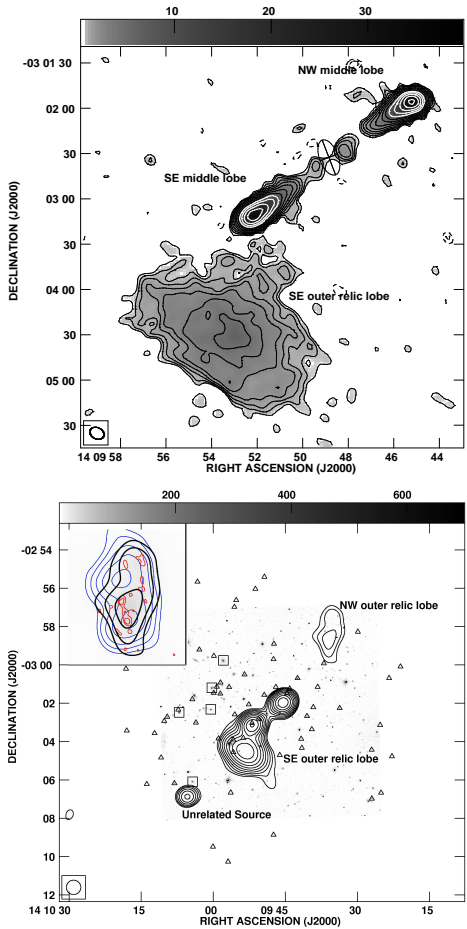,width=2.5in,angle=0}
  }
\caption[]{ Top Panel: The high resolution ($\sim$9 arcsec) GMRT 325
  MHz image, both in contour and grey-scale, of the inner, middle and
  SE outer relic lobes of Speca. The isophotal major-, minor-axis and
  P.A of the host galaxy, as in the text, are also indicated with an
  ellipse and '+' sign.  Bottom Panel: The low resolution (45 arcsec)
  GMRT 325 MHz image in contours is superposed on the $R$-band optical
  image. The relic lobes and an unrelated radio source are marked.
  The spectroscopically confirmed members of the cluster are marked by
  boxes, and possible members from SDSS $z_{\rm photo}$ are marked by
  triangles. The inset shows details of the NW outer relic lobe with
  NVSS data shown as blue contours, high resolution ($\sim$9 arcsec)
  GMRT 325 MHz image shown in red contours, and the low resolution (45
  arcsec) 325 MHz image as in the background is shown in thick black
  contours.}
\end{figure}
\label{optical}
\subsection{Disk structure in the host galaxy}
While comparing Sloan Digital Sky Survey (SDSS)
and Faint Images of the Radio Sky at Twenty-Centimeters (FIRST)
images, we identified the host galaxy Speca as an unusual
object. The galaxy lies at the geometrical centre of the radio lobes,
and the SDSS spectrum from the central region shows a 
[N{\small II}]/H$\alpha$ line ratio higher than unity, suggesting it to be
the AGN host galaxy. Our $R$-band image shows that the main body
clearly suggests a morphology consistent with a nearly edge-on disk
galaxy (Figure 1 (a)). The north-east (NE) side of the nucleus shows
a curved feature similar to a spiral-arm seen nearly edge-on.  Similar
structure is also seen in SDSS $g'$, $r'$ and $i'$-band images, but
any possible counterpart of this spiral arm on the south-western side
of the nucleus is not seen. The curved nature of the spiral arm-like
feature and extended isophotal contours on the southern side,
disfavours any mid-plane dust causing the observed stellar structure.
Measured up to 25 mag arcsec$^{-2}$ in the SDSS $r'$-band, the size of the
galaxy is 60 $\times$ 29 kpc (25 arcsec and 12 arcsec with
P.A. 21$^{\circ}$) suggesting that Speca is a large, mature host
galaxy.

In Figure 1 (b, c) we also present a comparison of the $R$-band image with
near-UV (NUV) and far-UV (FUV) images available on-line from the
Galaxy Evolution Explorer (GALEX). Surprisingly, the UV emission
(smoothed with a tophat kernel of radius 2 (4) pixels for NUV (FUV))
correlates nicely with the optical major-axis. The lack of dominant
UV emission from the centre of the galaxy suggests that the emission
is associated with star formation and unlikely to be dominated by the
central AGN.  In the NUV the northern side is brighter than the
southern side, consistent with the extinction being a possible cause
behind the lack of a spiral-arm feature on the southern side.
Additionally, we note that the optical colour is also redder for the
southern side.  UV images thus clearly demonstrate a young
star-forming disk in Speca.  From the UV color for the whole area of
Speca (FUV-NUV $\sim$ 0.43 in AB) and comparing with models of stellar
population synthesis (Bruzual \& Charlot 2003), we inferred that the
star formation in Speca is younger than $\sim$500 Myr. However, the
$u'$-$r'$ colour (2.6 in AB), five band SED and nuclear spectrum from
SDSS suggest a several Gyr old stellar population dominant in the host
galaxy.

The surface brightness profile of the Lulin $R$-band image was
analyzed with GALFIT version 2.03b (Peng et al. 2002). We explored
both the exponential disk and de Vaucouleurs profile, with and without
a nuclear Gaussian source. The best model was found to be an
exponential disk model (effective radius = 5.3 arcsec, axial ratio =
0.35, P.A.=17$^{\circ}$) with an unresolved (within the errors) point
source at the nucleus. The radial profile of the surface brightness is
shown in Figure 1 (d) with our best model and the next best
one (de Vacouleurs profile without a nuclear source). The observed
profile shows excess features on both the NE and SW sides of the
nucleus at $\sim$7 arcsec from the nucleus, making the fit
unsatisfactory in terms of statistics even with our best
model. However, the best model of disk with a nuclear point source is
far better than the second best one since the latter predicts too much
flux in both the inner and outer parts of the radial profile.

In summary, the optical and UV structure and optical radial brightness
profile strongly support the identification of Speca as a rare radio
galaxy hosted in a young star forming spiral/disk galaxy.  To the best
of our knowledge, this is the second spiral-host large radio galaxy,
after the only well-known case of 0313-192, discovered by Ledlow, Owen
\& Keel (1998).  In light of recent H{\sc i} studies finding large
massive gas disks (Oosterloo et al. 2007) and UV studies finding young
stellar disks in a large-fraction of early-type galaxies (Yi et
al. 2005, Salim \& Rich 2010), we speculate that the disk nature of
Speca is due to recent star formation, possibly following a merger
more than 500 Myr ago (e.g. NGC~3801; Hota et al. 2009, 2011).
\subsection{Episodic radio jet activity}
\label{radio}
The possible episodic nature of this radio galaxy was first identified
from comparisons of NVSS and FIRST images, where two prominent
emission blobs on large scales seen in NVSS were found to be missing
in FIRST. To investigate further we observed it with the GMRT at 325
MHz. We found typical `edge-brightened' radio lobe structure as in FR
II type radio galaxies (Figure 2 top panel).  A similar radio
structure is also seen in the higher frequency FIRST image, where the
lobes have equal flux densities of 24 mJy (23.9 mJy for north-western
(NW) and 23.8 mJy for south-eastern (SE)) within the errors. The
projected linear size of the lobes, 307 kpc (peak to peak measured to
be 128 arcsec aligned at P.A. of 306$^\circ$), is typical of classical
large radio galaxies. These components are hereafter referred to as
`middle lobes' of Speca.  To our surprise, we also found two new peaks
(2.24 mJy beam$^{-1}$ for NW peak and 1.34 mJy beam$^{-1}$ for SE
peak) of radio emission with a peak to peak projected separation of
$\sim$50 kpc (22 arcsec) at a P.A. of 297$^{\circ}$.  The total flux
densities of these blobs are 4.52 and 2.56 mJy for the NW and SE
respectively.  The position angle of the line joining these two peaks
is $\sim$10$^\circ$ offset from that defined by the middle
lobes. Higher resolution and higher frequency images are required to
confirm these peaks as radio lobes, but in comparison with B0925+420
(Brocksopp et al. 2007), we believe them to be lobes and refer as
`inner lobes'.  Interestingly, the inner lobes are located along the
minor axis of the host galaxy but are larger than its optical extent
(29 kpc) suggesting that the lobes have not formed due to interaction
with the inter-stellar medium of the host galaxy (Figure 2 top panel).

A large diffuse emission blob to the south of the SE middle lobe, and
almost connected, can be seen in Figure 2 (top panel).  It shows a
centrally peaked structure and the outer boundary on its SE side is
almost straight with a sharp drop in surface brightness.  The total
extent of this diffuse emission is $\sim$300 kpc, comparable to the
full extent of the middle lobes. It shows an overall NE-SW elongation
roughly orthogonal to the middle lobes. This emission blob was
prominent in NVSS but was missing in FIRST.  Because of this diffuse
nature and no apparent optical counterpart (see Section 3.3) hereafter
we refer to it as the `SE outer relic lobe'.

Corresponding emission on the NW side was very faint, and so we
present the smoothed 325 MHz image of a larger region of the sky in
Figure 2 (bottom panel).  In addition to the SE outer relic lobe and
middle lobes, an emission blob located NW of the middle lobes shows a
north-south elongation similar to that seen in NVSS.  Detailed
comparison of its structure from all the three images available (325
MHz with a beam of 9 arcsec (red contours), 325 MHz with a beam of 45
arcsec (thick black contours) and 1400 MHz with a beam of 45 arcsec
(blue contours)) is shown in the inset image. Although the overall
orientation matches, the location of the emission peaks does not
match, demonstrating the diffuse nature of the blob, and hence
explaining why it was missed in the high angular resolution (beam
$\sim$5 arcsec) FIRST data. This diffuse nature along with the lack of
any potential optical counterpart (see Section 3.3) suggests that it
is not an independent radio source.  Hereafter we refer to this as `NW
outer relic lobe'.

These relic lobes are quite asymmetric in terms of their location with
respect to the host galaxy, radio flux densities and orientation with
respect to the inner and middle lobes.  There is no signature of any
more diffuse emission apart from the lobes described here.  Note that
a compact radio source seen on the SE corner of the image is an
unrelated source with a clear optical counterpart.  The structure of
these two relic lobes neither bear similarity with typical radio lobes
nor with the partial shell-like structures of double radio relics
found in clusters (e.g. Bagchi et al. 2006).  Hence we believe that
both these diffuse radio blobs are due to an earlier episode of
jet-activity of the AGN hosted in Speca.  However, we can not
completely rule out an origin for this diffuse emission by some
independent radio-loud AGN or other cluster related phenomena.

The largest angular separation between these two relic lobes (570
arcsec) corresponds to $\sim$1.3 Mpc, classifying Speca as a giant
(old) radio galaxy (see Schoenmakers et al. 2001).  These three
distinct pairs of lobes are probably produced by three different
episodes of radio-jet activity of the AGN in Speca.  If confirmed,
this will be the second triple-double radio galaxy after B0925+420
(Brocksopp et al. 2007). Multiple episodes of jet activity may be
correlated with the recent ($\lapp$ 500 Myr) star formation seen in
the host galaxy (see Section 3.1).  Since these relic lobes could be
several 100s of Myr old and the few detections of them so far are
believed to be the ``tip of the iceberg'' (Dwarakanath \& Kale 2010),
upcoming sensitive low frequency surveys (e.g. TGSS: TIFR GMRT Sky
Survey at 150 MHz or surveys from LOFAR: Low Frequency Radio Array)
will possibly discover many more examples to help understand the duty cycle of
AGNs, black hole coalescence (Liu 2004), and the role of AGN-feedback
in galaxy and cluster evolution (Croton et al. 2006).
\subsection{Revived relic lobes in a cluster}
\label{cluster}
Speca belongs to a cluster of galaxies MaxBCG J212.45357-03.04237,
having 14 members within $r_{\rm 200}$ (Koester et al. 2007).  While
spectroscopic redshifts are available for only five other members in
NED (Figure 2 bottom panel), we searched the SDSS database for further
possible member galaxies with 0.135 $<$ $z_{\rm photo}$ $<$ 0.141 and
within a radius of 8 arcmin ($\sim$1 Mpc for the distance of
Speca). We found nearly 60 members (marked by triangles in Figure 2
bottom panel), further supporting the presence of a cluster of galaxies.
The maximum velocity difference of $\sim$1000 km s$^{-1}$ between
confirmed members also supports the presence of a cluster of galaxies
around Speca.  Although X-ray emission from this region has not been
detected by the ROSAT All Sky Survey, we calculate that a 3$\sigma$ upper
limit corresponds to a luminosity of $7 \times 10^{43}$ erg s$^{-1}$
and it is still consistent with the presence of a hot intra-cluster
medium around Speca.  There is no member galaxy located in the middle
of the diffuse emission.  Furthermore, by searching in the SDSS and
NED database without any redshift constraints, we do not see any
extended optical source ($>$2 arcsec) brighter than -18 in absolute
$r'$-band magnitude (assuming SDSS $z_{\rm photo}$) and potential
radio-loud AGN host galaxy (red in the colour magnitude diagram)
within the region of the diffuse lobes.

To understand the physical properties we investigated the spectral
index and polarisation of the relic lobes.  The two point spectral
index of the NW relic, which has $S_{325}$ = 46 mJy (from GMRT) and
$S_{1400}$ = 28.45 mJy (from NVSS), is very flat ($\alpha$=-0.32,
where $S_{\nu} \propto \nu^{\alpha}$).  There is no chance of missing
flux in our measurements at 325 MHz with the GMRT, for this small lobe
of nearly 3 arcmin size.  Although the spectrum of the relic lobe
should be confirmed by UV-matched imaging at multiple frequencies,
flattening is significant and unlikely have been caused by synthesis
imaging issues or the 5-10 per cent absolute flux density calibration
errors.  On the other hand if NVSS has missed flux, then the true
spectral index will be even flatter than -0.32.  So, although the
exact value of the spectral index ($\alpha$=-0.32) may not be
accurate, its true spectral index has to be, we believe, flatter than
typical cluster radio relics or relic lobes, which are usually steeper
than -1 (Ensslin et al. 1998, Dwarakanath \& Kale 2009).  Therefore,
we propose that such a flat spectral index of the NW relic lobe has
been caused by particle re-acceleration process (Brunetti et al. 2008,
van Weeren et al. 2010).

In NVSS, the SE middle lobe and the SE outer relic lobes are not well
separated.  Their total flux density is 108 mJy at 1400 MHz. However,
from the FIRST image we see only the SE middle lobe, and its flux
density is 24 mJy.  Subtracting this value from the total flux density
of the SE middle and the SE outer lobes we calculate the flux density
for the SE outer relic to be 84 mJy at 1400 MHz.  From the high
resolution GMRT measurements, the flux density of SE relic lobe at 325
MHz is 250 mJy.  The resulting two point spectral index of the SE
relic lobe is -0.75, which is $\sim$0.2 flatter than that of the
middle lobes (-0.92 for NE and -0.98 for SW lobes between 1400 MHz
(FIRST) and 325 MHz).  This spectral index of the SE relic is also
flatter than what is expected for such diffuse relic lobes at a
similar redshift (e.g. Dwarakanath \& Kale 2009).  So, some degree of
particle re-acceleration is also required in the SE relic lobe.  The
SE relic lobe shows a sharp linear edge and drops in surface
brightness on the edges.  The fraction of polarized flux density,
measured in NVSS, is 19 per cent, strongly suggesting ordering of the
magnetic field due to large-scale linear compression, like in shocks
(Ensslin et al. 1998, van Weeren et al. 2010).

The orientation of these elongated outer relic lobes shows little
relation with the jet direction of the middle lobes.  In the absence
of an X-ray image, the structure of the relics can be compared with
the galaxy distribution in the cluster.  All the five bright,
spectroscopically confirmed members are seen to the NE of Speca and
seem to be early-type galaxies based on their morphology and red
optical colour ($u'$-$r'$).  Speca is the largest and most dominant
member of the cluster but the distribution of all possible sixty
member galaxies does not show a higher concentration around it (Figure
2 bottom panel).  This distribution suggests Speca to be in a
dynamically young cluster, possibly related to either a merger or
accretion from galaxy filaments.  A cluster-cluster merger seems to be
unlikely a cause, as the distribution of member galaxies does not show
a double peak.  Furthermore, the morphology of the relic lobes does
not resemble a case where arc-like radio emission is seen
perpendicular to the elongated X-ray emission or double-peaked galaxy
distribution of the merging clusters (Bagchi et al, 2006, van Weeren
et al. 2010).  Any cluster-cluster merger along the line of sight can
also be ruled out since the relics are widely separated 
($\sim$1.3 Mpc) as seen in projection.  Therefore, we speculate that the
relic lobes are most plausibly revived by shocks at the cluster
outskirts due to cosmological accretion from galaxy filaments (Ensslin
et al. 1998; Pfrommer et al. 2008). Numerical simulations predict 
two locations of such accretion shocks, one near the virial radius (R$_{vir}$) 
and the other at $\sim$ 3R$_{vir}$ (Molnar et al. 2009 and references there in). 
Following the formula from Girardi et al. (1998), a rough estimate of the 
projected velocity dispersion ($\sim$ 300 km s$^{-1}$) of the spectroscopic 
members leads to a R$_{vir}$ of $\sim$ 850 kpc. Since re-accelerated relic
lobes can be expected at diametrically opposite ends of the cluster, the 
expected separation is $\sim$ 1.7 Mpc.  Given the unknown orientation of the 
relic lobes to the line of sight, the largest separation of 1.3 Mpc between 
the relic plasma is consistent with it tracing the accretion shock at 
the virial radius of this cluster.

Since the cluster is possibly dynamically young and Speca is the
brightest cluster galaxy (BCG) with young star formation, as argued in
similar BCGs (McNamara et al. 1996, Hicks, Mushotzky \& Donahue 2010),
we speculate that, alternatively to a galaxy merger origin, Speca
could be accreting cold gas from the surrounding intra-cluster
medium. Such young star-forming radio galaxies may be a `missing
link', as they are rare in the nearby Universe but likely common at
higher $z$ (Norris et al. 2007, Zirm et al. 2005).  Galaxy mergers
and/or cold gas accretion are likely to be correlated with the
multiple episodes of AGN jet activity we see in Speca.  Most UV-bright
young star formation seen in early-type galaxies is interpreted as
residual star formation or incomplete quenching of the star formation
by past AGN feedback (Schawinski et al. 2007). If this is the case,
then the multiple episodes of feedback seen in Speca with an unusual host
galaxy are a unique laboratory to investigate the galaxy-black hole
co-evolution process in action, and an opportunity for near-field
cosmology studies.
\section*{Acknowledgments} 
We are grateful to the referee, William Keel, for helpful comments. 
We thank the staff of the GMRT who made these observations
possible. GMRT is run by the National Centre for Radio Astrophysics of
the Tata Institute of Fundamental Research, India.  We thank the staff
and observers at Lulin Observatory, National Central University,
Taiwan. We are grateful to SDSS, GALEX, NVSS and FIRST surveys, and NED database.
AH is grateful to Yen-Ting Lin for his help in searching SDSS. SCR was
supported by the National Research Foundation of Korea to the Center
for Galaxy Evolution Research.

\end{document}